\documentclass[%
 aip,
 amsmath,amssymb,
reprint,%
]{revtex4-1}

\usepackage{graphicx}
\usepackage{dcolumn}
\usepackage{bm}

\usepackage[utf8]{inputenc}
\usepackage[T1]{fontenc}
\usepackage{mathptmx}
\usepackage{etoolbox}

\makeatletter
\def\@email#1#2{%
 \endgroup
 \patchcmd{\titleblock@produce}
  {\frontmatter@RRAPformat}
  {\frontmatter@RRAPformat{\produce@RRAP{*#1\href{mailto:#2}{#2}}}\frontmatter@RRAPformat}
  {}{}
}%
\makeatother

\newcommand{\siformat}[2]{${#1}\,\mathrm{#2}$}

\begin{document}

\preprint{AIP/123-QED}

\title{A Superconducting Nanowire Binary Shift Register}
\author{Reed A. Foster}
\email{reedf@mit.edu}
\author{Matteo Castellani}
\author{Alessandro Buzzi}
\author{Owen Medeiros}
\author{Marco Colangelo}
\author{\\Karl K. Berggren}
\affiliation{Department of Electrical Engineering and Computer Science, Massachusetts Institute of Technology, Cambridge, Massachusetts 02139, USA}

\date{\today}

\begin{abstract}
  We present a design for a superconducting nanowire binary shift register, which stores digital states in the form of circulating supercurrents in high-kinetic-inductance loops.
  Adjacent superconducting loops are connected with nanocryotrons, three terminal electrothermal switches, and fed with an alternating two-phase clock to synchronously transfer the digital state between the loops.
  A two-loop serial-input shift register was fabricated with thin-film NbN and achieved a bit error rate less than 10$^{-4}$, operating at a maximum clock frequency of \siformat{83}{MHz} and in an out-of-plane magnetic field up to \siformat{6}{mT}.
  A shift register based on this technology offers an integrated solution for low-power readout of superconducting nanowire single photon detector arrays, and is capable of interfacing directly with room-temperature electronics and operating unshielded in high magnetic field environments.
\end{abstract}

\maketitle


Superconducting nanowires are interesting candidates for cryogenic data processing and storage, particularly for readout of superconducting nanowire single photon detector (SNSPD) arrays.
The high kinetic inductance of thin film superconductors allows them to store data in compact loops,\cite{ZhaoYTronHTronMemory} and the existence of nanocryotrons (nTrons), three-terminal electrothermal switches, \cite{nTron} enables the creation of low-power digital logic and memory elements.\cite{BuzziNTronLogic}
In addition, superconducting nanowires can operate in harsh environments.
NbN is radiation hard,\cite{NbNNeutronHard} and SNSPDs have been shown to operate under high magnetic fields: both in-plane up to \siformat{5}{T} and out-of-plane up to \siformat{500}{mT}.\cite{SNSPDsInBField}
This makes nanowires an interesting candidate for applications in which SNSPD readout electronics must be able to withstand strong ambient magnetic fields or radiation, such as high energy physics and space exploration.
Furthermore, the shared technology platform with SNSPDs and ability to drive high-impedance loads\cite{nTron} is a strong motivator for direct integration of nanowire electronics with SNSPD arrays.
Dedicated readout electronics are necessary to address the thermal and mechanical challenges of scaling SNSPD imagers beyond 1 kilopixel, \cite{SNSPDArrayCryoLoad} and low-power electronic devices that operate in extreme environments and can be fabricated adjacent to superconducting detectors are an attractive choice over Josephson junction logic and CMOS.
Previous work \cite{ZhaoDelayLine,ThermallyCoupledImager} has used the high kinetic inductance of superconducting nanowires to make analog delay-line imagers, which offer high pixel counts and preserve the picosecond timing resolution of the SNSPDs.
Row column multiplexing has also been shown as an effective technique for reducing cable counts, \cite{RowColumnImager} however a more aggressive reduction in cable count will be required for megapixel arrays.
Inspired by the operation of a semiconductor CCD, serial readout of SNSPD arrays could be performed by a superconducting nanowire binary shift register.
Serial readout may enable higher count rates than delay-line techniques by shortening dead-time, but more importantly, it simplifies the interface to conventional CMOS readout electronics by removing the need for high resolution, low jitter time-to-digital converters.

\begin{figure*}[htbp]
  \includegraphics{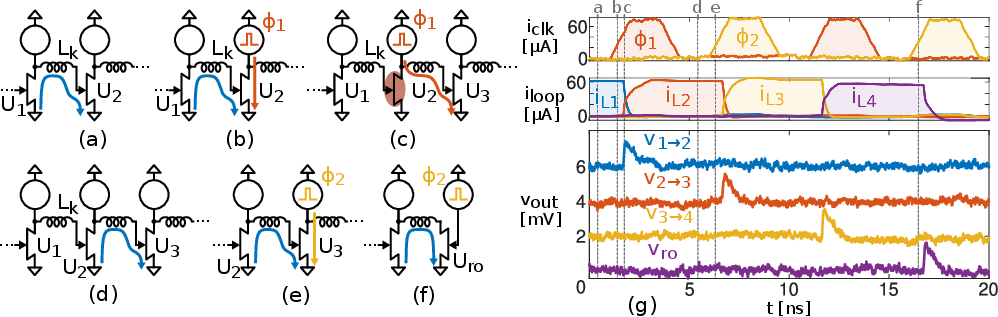}
  \caption{
    (a)-(f) Principle of operation of the shift register, which uses the presence or absence of circulating current to encode digital states.
    (g) Shows the results of a transient simulation in LTSpice of a four-loop shift register, including noise and parasitics from the packaging and experimental apparatus.
    (a) Shows a shift register with an initial circulating current in the loop formed by the kinetic inductor L$_\text{k}$ and nTrons U$_1$, U$_2$.
    The corresponding time $a$ in the simulation is indicated in (g).
    A two phase clock ($\phi_1$, $\phi_2$) is used to transfer the digital state between adjacent loops; the first phase $\phi_1$ is applied in (b).
    In (c), the summation of the clock and circulating currents exceeds the switching current of U$_2$'s channel, forming a resistive hotspot and diverting the clock into the loop formed by U$_2$ and U$_3$.
    The hotspot creates a voltage spike $v_{1\rightarrow2}$ shown in the lower panel of (g) at time $c$.
    By the time the clock is turned off in (d), the channel of U$_2$ has healed and a circulating current is present in the loop between U$_2$ and U$_3$.
    The process continues in (e) when the second clock phase $\phi_2$ is applied.
    Two clock phases are needed to ensure a zero resistance path to ground for the diverted clock, for example, the path through U$_3$ as shown in (c).
    The readout nTron U$_\text{ro}$ in (f) is used to reset the state of the final loop and generate an output voltage conditional on the presence of a circulating current.
  }
  \label{fig:shiftreg}
\end{figure*}

In this work, we demonstrate a proof-of-concept for a superconducting nanowire binary shift register, which encodes digital states with dissipationless circulating current in superconducting loops.
As shown in Fig. 1, each loop is formed by a kinetic inductor L$_\text{k}$ and two nTrons, U$_1$ and U$_2$.
The presence of a circulating current flowing through $L_\text{k}$ into the gate of U$_2$ encodes a binary ``1'', and the absence of current is used to represent a ``0''.
The shift register is designed to use circulating currents on the order of \siformat{100}{\mu A}, therefore small ($\mu A$) fluctuations in loop current (\textit{e.g.} due to thermally-activated phase slips which change the stored flux in each loop by $\Phi_0$) are not expected to impact the binary state.
A substantial environmental disturbance that makes the film resistive (e.g. $T > T_c$, $H > H_{c2}$) would be necessary to destroy the state stored in the shift register.
The state of the shift register is only altered under the application of a clock, when the combination of the circulating current and clock pulse exceeds the critical current density in the nTron channel, causing it to switch from superconductive to resistive, diverting the clock pulse into the next loop.
This process forms a new circulating current conditional on the presence of current in the previous loop.
A two-phase clock is used to guarantee the diverted current always has a superconducting path to ground (as shown in Fig. \ref{fig:shiftreg}c).
In comparison to the original nTron design,\cite{nTron} which acts like an amplifier, the loops are connected with wide-gate nTrons, where the width of the gate constriction is comparable or equal to that of the channel constriction.
A wide-gate nTron is crucial for the shift register: because the output of one nTron becomes the input of another, the current levels for the input and output should be equal.
The additional readout nTron shown in Fig. \ref{fig:shiftreg}f uses a standard nTron with a small choke.
It terminates the final loop of the shift register to destroy any circulating current present at the end of each clock cycle.
The readout nTron serves two purposes: (1) to reset the final loop of the shift register, and (2) to generate an output voltage signal, which can be sent to off-chip readout electronics or cascaded through a resistor to other nTron logic.

\begin{figure*}[htbp]
  \includegraphics{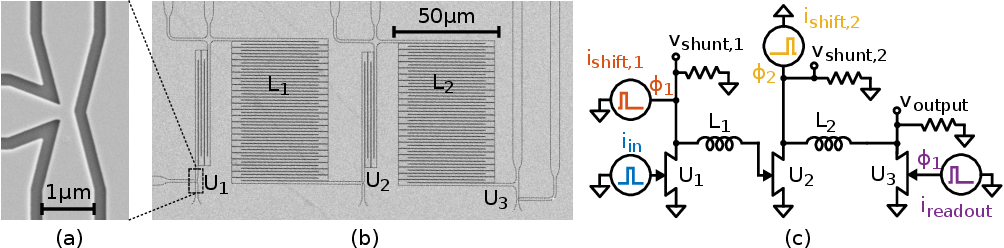}
  \caption{
    (a) and (b) show electron micrographs of fabricated wide nTron and two loop shift register.
    The large meanders in (b) are \siformat{100}{nH} kinetic inductors.
    (c) Is an equivalent circuit model of the experimental circuit.
    The current pulses are provided by a voltage source in series with \siformat{2}{k\Omega} resistors mounted off-chip on a printed circuit board.
  }
  \label{fig:setup}
\end{figure*}

The shift register was fabricated on a \siformat{16}{nm}-thick layer of NbN, deposited with an AJA sputtering system onto an Si wafer with \siformat{300}{nm}-thick SiO$_2$ thermal oxide.
The circuit geometry was patterned on the NbN layer with electron-beam lithography using ZEP530A resist and CF$_4$ reactive ion etching.
The wide-gate nTron channel constriction widths were designed to be \siformat{270}{nm} (with an equal-sized gate choke), and the readout nTron channel width was designed to be \siformat{240}{nm}, with a gate choke width of \siformat{40}{nm}.
Figure \ref{fig:setup}a shows an electron micrograph of a wide-gate nTron patterned on thin-film NbN.
Figure \ref{fig:setup}b is an electron micrograph of the experimental two-loop shift register circuit, and the equivalent circuit model is shown in Fig. \ref{fig:setup}c.
The loop kinetic inductors were designed to be \siformat{100}{nH}; the estimated inductance came out to \siformat{60}{nH} (\siformat{30}{pH} per square) based on a room temperature sheet resistance measurement of \siformat{194}{\Omega} per square.
The finished chip was wirebonded to a printed circuit board with off-chip current bias and shunt resistors, which was mounted to a custom dip probe\cite{BrendenThesis} and cooled to \siformat{4.2}{K} in a dewar of liquid helium.
The \siformat{2}{k\Omega} bias resistors were used as approximate current sources to convert an applied voltage to a current through the nanowire.
The hotspot resistance of the switching nTron is small compared to \siformat{2}{k\Omega}, so the amount of current through the nanowire given some applied voltage stays roughly constant regardless of the nanowire state.
The nTron dimensions, inductor sizes and resistor values were selected through LTSpice simulation,\cite{BerggrenSpice} the results of which are shown in Fig. \ref{fig:shiftreg}g.
The bit error rate of the shift register model under high levels of noise (\textit{e.g.} $\pm5\%$ variation in clock amplitude) was used to guide selection of component properties.
Eight different shift register circuits were fabricated on a single \siformat{1}{cm^2} chip.
Two circuits were tested: the circuit presented in this letter, which used a wide-gate nTron to connect adjacent loops, and a shift register with a different switch geometry.
The alternative design used current summation into a single two-terminal constriction as a switch, which performed worse than the design based on the wide-gate nTron, likely due to leakage current that could flow between loops unimpeded regardless of the switch state.
The results presented in this letter are from the circuit which used wide-gate nTrons.

\begin{figure*}[htbp]
  \includegraphics{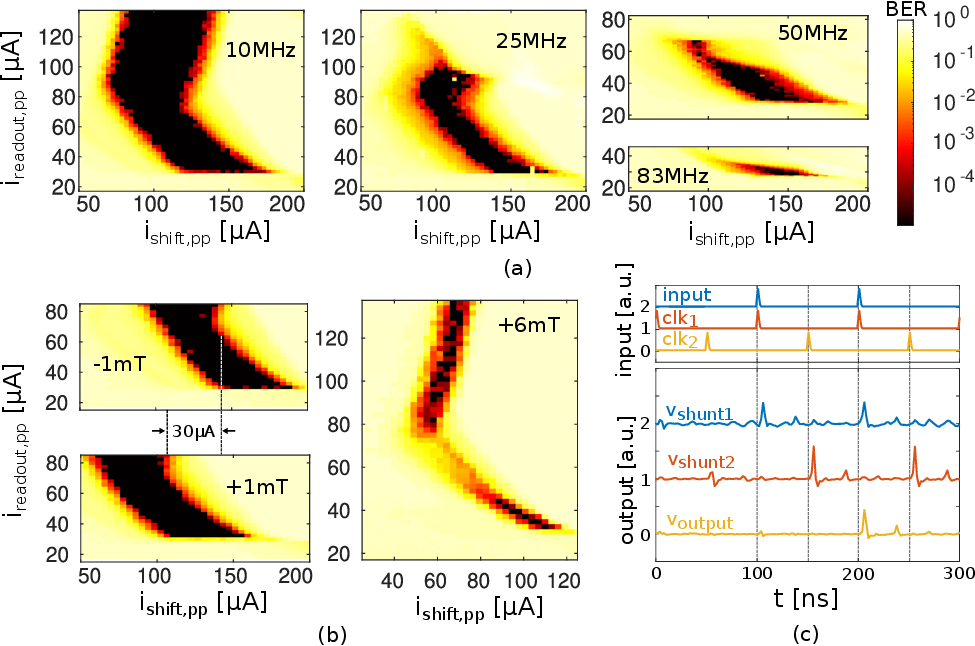}
  \caption{
    (a) and (b) are bias margin plots, which show the bit error rate of the shift register (number of errors out of a \siformat{10}{kbit} random bit sequence) as a function of shift and readout clock amplitude. The black regions represent correct operation, with a bit error rate below $10^{-4}$.
    The input clock amplitude was fixed at a level that gave optimal margins.
    (c) Shows an example trace of the transient response of the circuit with a \siformat{10}{MHz} clock.
    The voltage spikes on V$_\text{shunt1}$ indicate the storage of a circulating current in the first loop, and spikes on V$_\text{shunt2}$ indicate transfer of state between adjacent loops.
    Traces are vertically offset for clarity.
  }
  \label{fig:results}
\end{figure*}

The circuit was characterized with clock rates from \siformat{10}MHz to \siformat{100}{MHz} and under magnetic fields from \siformat{\pm1}{mT} to \siformat{\pm6}{mT}, applied orthogonal to the chip surface by a superconducting magnet mounted on the end of the dip probe.
A Keysight PXIe M3202A (arbitrary waveform generator) and M3102A (digitizer) were used to verify correct operation of the shift register over a range of signal amplitudes.
This was done by generating multiple \siformat{10}{kbit}-long pseudorandom binary sequences of voltage pulses and measuring the circuit response.
The data and clock input signals encoded digital ``1''s with low-duty-cycle \siformat{2}{ns} FWHM voltage pulses, as can be seen in the top panel of Fig. \ref{fig:results}c.
The PXIe chassis controller swept the amplitude of the shift and readout clock pulses and measured the bit error rate in near real time for each set of clock amplitudes by comparing the device output with the \siformat{10}{kbit} input sequence.
Each spike of the output waveform was thresholded and digitized, and the result was compared with a copy of the input signal delayed by a clock period --- for each instance where the input and digitized output differed, the total error count was incremented.
A sample waveform used to calculate the bit error rate is shown in Fig. \ref{fig:results}c.

The plots in Fig. \ref{fig:results}a are bias margin plots, which show the bit error rate as a function of clock pulse amplitude for various clock rates.
The dark regions indicate no measured errors for the \siformat{10}{kbit} sequence, and the width of the dark regions give the bias margins, defined as the amount of variation in clock amplitude that is acceptable before the circuit begins to function incorrectly.
The device performed correctly up to a maximum clock rate of \siformat{83}{MHz}, with the bias margins steadily shrinking for increasing clock frequency.
The bias margins of the shift clock were $\pm$\siformat{24}{\%} at $f_\text{clk} = 10$\siformat{}{MHz}, but only $\pm$\siformat{7}{\%} for $f_\text{clk} = 83$\siformat{}{MHz}.
Margins for the readout clock shrank even more, from $>\pm$\siformat{45}{\%} at $f_\text{clk} = 10$\siformat{}{MHz} to $\pm$\siformat{5}{\%} at $f_\text{clk} = 83$\siformat{}{MHz}.
As shown in Fig. \ref{fig:results}b, the introduction of a $\pm$\siformat{1}{mT} field did not dramatically hurt the margins of the shift clock: $\pm$\siformat{25}{\%} for +\siformat{1}{mT} and $\pm$\siformat{20}{\%} for -\siformat{1}{mT}.
The readout clock margins were unimpacted.
However, introduction of a +\siformat{6}{mT} field reduced the margins of the shift clock to $\pm$\siformat{4}{\%}, and a -\siformat{6}{mT} field (not shown) prevented the device from working with a bit error rate below 10$^{-3}$.

The lower half of each bias margin plot exhibits a downwards slope due to the transfer characteristics of the readout nTron: for a larger gate current, the required channel current to switch the nTron is lower.
Therefore, for a larger readout clock, the required loop current (and thus shift clock amplitude) is lower.
The abrupt change in bit error rate for readout clock amplitudes below \siformat{30}{\mu A} occurred because the readout clock was not strong enough to switch the readout nTron.
If the final loop current is left circulating, it prevents the middle nTron from switching again when a shift clock is applied.
The optimal bias region slopes upwards for high readout clock currents, possibly because of current injection from the readout clock, which would create a reverse circulating current in the final shift register loop.
This would require the amplitude of the shift clock to be larger to leave a net-forward circulating current in the final loop that was large enough for the readout nTron to switch when clocked.

As the frequency of the clock increased, the bias margins for the shift clock shrank from both sides, and the maximum acceptable readout clock amplitude decreased dramatically.
The $L/R$ time constant to charge a loop with a circulating current depends on the loop kinetic inductance and the total shunt resistance.
It is plausible that, for higher clock frequencies, the circulating current does not reach a stable level in the half-period between the two clock phases, thus producing incorrect behavior.
Further characterization with various shunt resistor and kinetic inductor sizes should be performed to verify that the decrease in margins is due to this electrical time constant, and not a thermal process or some other unconsidered effect.
One possible explanation for the large decrease in the bias margins of the readout clock could be slow thermal reset of the readout nTron gate choke.
The designed critical current of the choke was only \siformat{30}{\mu A},
and overdriving the readout clock significantly above that (e.g., \siformat{100}{\mu A}) would generate a considerable amount of heat.
Residual heat from a readout clock with phase $\phi_1$ would suppress the critical current of the channel, potentially causing the readout nTron to switch on phase $\phi_2$ if it had not cooled sufficiently.
Shunting the gate with a small resistor could limit the heating of the choke, potentially restoring the bias margin range of the readout clock for high clock frequencies.

The observed shift in bias margins of \siformat{15}{\mu A/mT} due to the external magnetic field (Fig. \ref{fig:results}b) agrees with the expected loop current induced by the Meissner effect.
However, enhancement of current crowding around constrictions (such as the sharp corners in the nTron channel as can be seen in Fig. \ref{fig:setup}a) due to the Lorentz force is potentially a more plausible explanation, so further work must be done to understand the mechanism of the external field on the bias margins of the circuit.
If the Meissner effect is the dominant mechanism, reducing the size of the loop inductor may help improve resilience against out-of-plane magnetic fields.
Instead, if the mechanism is current crowding enhanced by the Lorentz force, then the nTron geometry would need to be modified to mitigate this effect.

The total energy of any cryogenic electronics system will be dominated by the cryocooler, which can consume on the order of \siformat{1}{kW} to supply tens of milliwatts of cooling power at \siformat{4}{K}.\cite{SmallCryocoolers}
Unless the design of the shift register presented in this work is modified, SNSPD arrays using shift register readout are limited to the kilopixel regime by cryostat cooling power.
The energy consumption of the shift register is estimated to be \siformat{80}{fJ} per shift operation, and is dominated by the clocking: each clock phase dissipates \siformat{100}{\mu A} through \siformat{2}{k\Omega} for \siformat{2}{ns}.
When the shift register stores a ``1'', approximately \siformat{300}{aJ} of energy is stored (\siformat{100}{\mu A} in a \siformat{60}{nH} loop).
Each shifting operation destroys this circulating current, dissipating the stored energy through the resistive hotspot in the nTron channel.
Shift register readout of a 1 kilopixel array clocked at \siformat{50}{MHz} would dissipate about \siformat{4}{mW}.
Reduction of the clock impedance by a factor of 20 from \siformat{2}{k\Omega} to \siformat{100}{\Omega} and the operating current from \siformat{100}{\mu A} to \siformat{10}{\mu A} would reduce the power dissipation of the 1 kilopixel array to \siformat{2}{\mu W}, making a megapixel array feasible from a power perspective.

Decreasing the size of the loop inductor will enable faster, more compact shift registers due to a reduced kinetic inductance and therefore smaller $L/R$ loop current time constant.
The speed of the device is fundamentally limited by the hotspot thermal relaxation time, since the nTron channel must cool between the two clock phases, otherwise there will not be a superconducting path for the diverted clock if the previous shift register stage switches. For example, as shown in Fig. \ref{fig:shiftreg}c, U$_3$ must be superconducting during the application of clock $\phi_1$.
An nTron fabricated with NbN on SiO$_2$ thermal oxide has achieved a thermally-limited switching speed of \siformat{615.4}{MHz}, with an estimated thermal relaxation time of \siformat{130}{ps}.\cite{nTronSwitchingSpeed}
Based on this, a conservative estimate for the thermal-reset-limited clock frequency of the shift register is about \siformat{1}{ns}, allowing for a \siformat{500}{MHz} two-phase clock.
At this clock rate, a 1 megapixel array could be read out on two wires at a frame rate of \siformat{1}{kHz}, for a maximum photon count rate of \siformat{1}{Gcps}.
More thermally conductive substrates can speed up thermal relaxation, \cite{NbNHotspotRelaxation} potentially offering further speed improvements to nanowire logic.

Due to the small feature size of the nTron constriction, fabrication variations may pose a challenge when drastically reducing feature sizes, especially for shift registers with many nTrons.
In order to minimize cable count, the same clock signal must be shared between multiple nTrons for any practical shift register.
Therefore, all nTrons will receive the same amplitude clock signal, so if there is substantial variation in the switching current of the nTrons, then some loops may not function correctly for a clock amplitude which works for other loops.
The bias margins of each nTron in a large shift register will have roughly the same shape, with variations in the midpoint of the optimal bias region due to edge roughness altering the constriction widths.
Film thickness also plays a role, but edge roughness should be the dominant factor in switching current variations.
Based on Fig. \ref{fig:results}a, the allowable variation in switching current is $\pm$\siformat{7}{\%} for a clock rate of \siformat{83}{MHz}.
This is equivalent to $\pm$\siformat{18}{nm} variation in nTron width for the \siformat{270}{nm}-wide nTrons.
A nanowire fabrication process using ma-N demonstrated 36 nTrons with a mean gate width of \siformat{33.7}{nm} and standard deviation of \siformat{2.4}{nm} across a \siformat{1}{cm^2} chip area.\cite{maNnTronSwitchingCurrent}
With $\pm$7 standard deviations of allowable variation in width, a shift register with millions of nTrons should be feasible.
However, scaling down to smaller nTron widths may still pose a challenge, as the relative variation in nTron switching current is larger.

\begin{figure}[htbp]
  \includegraphics{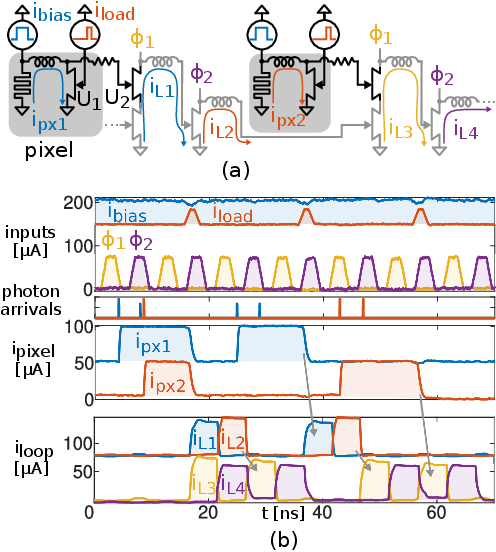}
  \caption{
    Proposed parallel readout scheme and transient simulation results.
    (a) Shows modifications to the original shift register (gray) in black.
    Each pixel consists of an inductively-shunted SNSPD which is read out with an nTron.
    When the bias $i_\text{bias}$ is enabled, photon arrivals divert the bias current into the right branch of the pixel.
    An additional bias current is applied to the $\phi_1$ clock input of each shift register stage.
    If the SNSPD bias current is diverted, a pulse of current $i_\text{load}$ applied to the gate of nTron U$_1$ will cause it to switch, sending a pulse of current to the gate of nTron U$_2$.
    The additional bias on the $\phi_1$ input will be diverted, forming a circulating current in the shift register.
    In (b), photon arrivals create circulating currents in each pixel.
    After the application of the load pulse $i_\text{load}$, the pixel states are loaded into the shift register and shifted out.
  }
  \label{fig:parallelreadout}
\end{figure}

Because the device we fabricated only accepts serial inputs, it would provide little practical benefit for large SNSPD arrays, as it is incapable of reducing wire count.
However, modifications to the circuit design can be incorporated to load data from an entire row of pixels in parallel into the shift register, as shown in Fig. \ref{fig:parallelreadout}.
This proposed modification was designed and simulated in LTSpice.
A simple pixel and destructive-readout memory can be implemented with an inductively-shunted SNSPD and nTron.
A second nTron is used to store a current in the shift register when the pixel is read out, conditional on the presence of a current in the pixel inductor.
Using this technique, data from all pixels could be loaded simultaneously into the shift register.
Since the readout of the pixels is destructive, the bias current through the SNSPD is restored, so the pixels can still detect photons after the pixel data is loaded into the shift register.
There is still per-pixel dead time set by the frame rate of the imager, since each pixel can only detect a single photon before it is reset again, but there is no imager-wide dead time like in a delay-line readout approach.

In addition to performing detector readout, the simplicity of a shift register makes it a useful test structure, which could be used to characterize process yield, as has been done in the past with SFQ logic to evaluate yield for Josephson junction processes. \cite{ShiftRegProcessVehicle}
More generally, the inherent ability of shift registers to serialize and deserialize data makes them a critical function of any large-scale digital system.
A superconducting shift register could help increase the capacity of links between room temperature and superconducting electronics, and with the introduction of digital logic, push even more computing into the fridge and enable larger scale superconducting systems based on nanowires.

The initial stages of this work were sponsored by the Army Research Office (ARO) under Cooperative Agreement Number W911NF-21-2-0041. The views and conclusions contained in this document are those of the authors and should not be interpreted as representing the official policies, either expressed or implied, of the Army Research Office or the U.S. Government. The U.S. Government is authorized to reproduce and distribute reprints for Government purposes notwithstanding any copyright notation herein.
The completion of the data analysis and presentation was funded by the DOE under the National Laboratory LAB 21-2491 Microelectronics grant.
The authors would like to thank Kyle Richards and Teja Kothamasu for assistance with setting up and using the Keysight PXIe system.
The data that support the findings of this study are available from the corresponding author upon reasonable request.
The authors have no conflicts of interest to report.

\bibliography{reedf_shiftreg}

\end{document}